\documentstyle [12pt] {article}

\newcommand{\cs}[3]{{{#3} \brace {#1 #2}}}

\newcommand{\h}[1]{\mathop{\lambda}\limits_{#1}\ \!\!\!}

\newcommand{\edf}{\ {\mathop{=}\limits^{\rm def}}\ }

\begin{document}
\begin{center}
\bf { THE ACCELERATING EXPANSION OF THE UNIVERSE AND TORSION
ENERGY  }
\begin{center}
{M.I.WANAS}\\ {\it Astronomy Department, Faculty of Science, Cairo
University, Giza, Egypt.\\
E-mail:wanas@mailer.eun.eg}
\end{center}
\begin{abstract}
In the present work, it is shown that the problem of the
accelerating expansion of the Universe can be directly solved by
applying Einstein geometrization philosophy in a wider geometry.
The geometric structure used to fulfil the aim of the work is a
version of Absolute Parallelism geometry in which curvature and
torsion are simultaneously non vanishing objects. It is shown
that, while the energy corresponding to the curvature of space-
time gives rise to an attractive force, the energy corresponding
to the torsion indicates the presence of a repulsive force. A fine
tuning parameter can be adjusted to give the observed
phenomena.
\end{abstract}
\end{center}
\section {Introduction}
 Recently, observation of red-shift of type $Ia$ supernovae$^{1}$
indicate that our Universe is in an accelerating expansion phase.
The four known fundamental interactions, as we understand them
today, can not account for the accelerating expansion of the
universe. Weak and strong interactions can be easily ruled out,
since both are very short range and cannot play any role in the
large scale behavior of our Universe. Also, electromagnetism
cannot play any role, in this respect, since neither the Universe
nor its material constituents are electrically charged, while
monopoles are not present in our Universe. The only interaction
which is playing the main role, in the structure and evolution of
the universe, is gravity. But gravity, as far as we know, is
attractive and not repulsive. Consequently, it can not give any
satisfactory interpretation of the accelerating phase of the
Universe. One can deduce that, either gravity is not well
understood and an important ingredient is missing in gravity
theories, or a new interaction is about to be discovered, in order
to account for such observations.

Gravity theories are usually constructed to give a better
understanding of gravitational interactions, from both
quantitative and qualitative points of view. Newton's theory deals
with gravity as a force acting at a distance. This theory has
suffered from some problems when applied to the motion in the
solar system (the advance of perihelion of Mercury's orbit). In
addition, the theory has shown its non-invariance under Lorentz
transformations. These problems motivated Einstein to construct a
new theory for gravity, the {\it "General theory of Relativity"}
(GR). In addition to the solution of the problems of Newton's
theory, GR gives very successful interpretation of gravitational
phenomena in the solar system, binary star system and many other
systems in the Universe. Also, GR has predicted the existence of a
number of phenomenae which has been confirmed observationally,
afterwards. Both Newton's theory and Einstein's theory give rise
to an attractive force. So, neither of these theories, in their
orthodox forms, can account for supernovae type Ia observations,
since this needs the existence of repulsive force which is missing
in both theories.

Now since the problem of SN type Ia observation has no
satisfactory solution and is still challenging , one cannot reject
any assumption which may lead to a solution. Among these
assumptions are:

(a) The existing physics is not adequate for interpreting all
known phenomena in our

Universe. One may need new physics which would imply new
interactions.

(b) The existence of a new type of matter in the Universe, with
new properties, that

can lead to the observed accelerating expansion of the Universe.

(c) One or more of the known fundamental interactions are not well
understood and

need further investigation. In particular, there may be a gap in
our present

understanding of gravity$^{1}$.

Many authors, in order to solve the above mentioned problem, have
returned back to a modified version of GR, in which the
cosmological constant existed in the theory (see Ref. 2).Although
this constant can solve the problem, since it may give rise to a
repulsive force, it suffers from a big problem the {\it "
cosmological constant problem" }$^{3}$. However, GR still has its
attractive feature i.e. its philosophical basis, the {\it "
geometrization philosophy"}. This philosophy still deserves
further investigations. It may provide solutions to gravity
problems especially those connected to SN observations.

The aim of the present work is to re-examine the geometrization
philosophy, seeking a solution for gravity problems. In the next
section, a brief review of the geometerization philosophy is
given. In section 3, a summary of the geometric structure, used in
the the present work, is given. Section 4 gives an attempt to
solve the problem addressed above using the torsion energy. The
work is discussed and concluded in section 5.
\section{Summary of the Geometrization Philosophy}
In constructing his theory of GR, Einstein has invented an
ingenious idea that geometry can be used to solve physical
problems. This idea is known as the {\it " Geometrization of
Physics"}. It comprises a philosophical principle which can be
summarized in the following statement.

{\underline{\it "To understand Nature one has to start with
Geometry and end
with Physics "}}.\\
Einstein has applied this philosophy using the following guide
lines: \\
1. Laws of nature are identities in the chosen geometry. \\
2. There is a one-to-one correspondence between physical
quantities and geometric

objects.
 \\
3. Physical trajectories of test particles are geometric paths
(curves) in the chosen

geometry. \\
Einstein has chosen the 4-dimensional Riemannian geometry, with
the following identifications corresponding to the 3-guide lines
given above$^{4}$, to represent the physical world
including space and time: \\
1. Conservation, as a law of nature, corresponds to the second
contracted Bianchi identity,
$$
 (R^{\mu \nu}- \frac{1}{2}g^{\mu \nu}R);_{\nu} \equiv 0   \eqno{(1)}
$$

where $R^{\mu \nu}$ is Ricci tensor, $R$ is Ricci scalar and the
semicolon is used for covariant

differentiation. \\
2. The conserved quantity (the quantity between brackets) ,
$$
G^{\mu \nu} \edf R^{\mu \nu}- \frac{1}{2} g^{\mu \nu}R, \eqno{(2)}
$$

is constructed from the curvature tensor (built using Christoffel
symbol). This

quantity corresponds to a type of energy that causes the curvature
of the space. {\it (For

this reason,  geometrically speaking, in what follows we are going
to call the quantity

defined by (2) {\bf "The
Curvature Energy")}}.\\
3. Taking the third guide line into consideration, Einstein has
used the geodesic equation
$$
\frac{d^{2}x^{\mu}}{ds^{2}} +
\cs{\alpha}{\beta}{\mu}\frac{dx^{\alpha}}{ds}
\frac{dx^{\beta}}{ds} =0, \eqno{(3)}
$$

to represent the trajectory of any test particle in the field and
the null-geodesic

equation,
$$
\frac{d^{2}x^{\mu}}{d\lambda^{2}} +
\cs{\alpha}{\beta}{\mu}\frac{dx^{\mu}}{d\lambda}
\frac{dx^{\nu}}{d\lambda} =0, \eqno{(3)}
$$

to represent the trajectory of photons.\\ On the above scheme, we
have the following comments: \\
1. The choice of Riemannian geometry represents a limited case,
since this geometry has

a vanishing torsion. This choice would not give a complete
description of the

physical world including space-time. Einstein has realized this
fact in his

subsequent attempts to construct unified field theories$^{4, 5}$
and has used

geometries with torsion in these attempts. \\
2. The use of equation (3)or (4) to describe motion of test
particles implies the application

of the equivalence principle. This may represent a desirable
feature in describing the

motion of a scalar test particle (particle defined completely by
its mass (energy)).

Note that, most of elementary constituents of the Universe are
fermions (particles

with mass, spin, charge, .... ), for which (3) and (4)
are no longer applicable. \\
3. The most successful results of GR, which confirm the theory,
are those obtained using

the field equations for empty space (pure geometric). The use of
the full field

equations of GR (equations containing a phenomenological matter tensor) is almost

problematic. \\
Considering these comments, we are going to examine the
application of the geometrization philosophy to geometries with
simultaneously non-vanishing  curvature and torsion.
 \section{Geometries with  Curvature and Torsion }
Torsion tensor is the antisymmetric part of any non-symmetric
linear connection. Einstein has used two types of geometries, with
non-vanishing torsion, in his attempts unify gravity and
electromagnetism. The first type is the Absolute Parallelism (AP)
geometry with non vanishing torsion but vanishing curvature$^{5}$.
The second  is of Riemann-Cartan type with simultaneously
non-vanishing torsion and curvature$^{4}$. In what follows, we are
going to review very briefly a version of AP-geometry in which
both torsion and curvature are simultaneously non-vanishing. We
have chosen this type since calculations in its context are very
easy.

The structure of a 4-dimensional AP-space is defined completely by
a set of 4-contravariant linearly independent vector fields
$\h{i}^{\mu}$. The covariant components of such vector fields are
defined such that (for further details see Ref. 6),
$$\h{i}^{\mu}\h{i}_{\nu}= \delta^{\mu}_{\nu}, \eqno{(5)}$$
$$\h{i}^{\mu}\h{j}_{\mu}= \delta_{ij} \eqno{(6)}.$$
The second order symmetric tensors,
$$g_{\mu \nu}  \edf \h{i}_{\mu}\h{i}_{\nu}, \eqno{(7)}$$
$$g^{\mu \nu}  \edf \h{i}^{\mu}\h{i}^{\nu}, \eqno{(8)}$$
can play the role of the metric tensor, and its conjugate, of the
Riemannian space associated with the AP-space. With the help of
(7) and (8) one can define a linear connection (Christoffel symbol
of the second kind), using which one can perform covariant
derivatives, as usual. Another linear connection can be obtained
as a consequence of AP-condition,
$$
\h{i}_ {\stackrel{\mu}{+} | \nu}= 0. \eqno{(9)}
$$
This equation can be solved to give the following non-symmetric
linear connection $\Gamma^{\alpha}_{. \mu \nu}$ (the (+)sign is
used to characterize covariant derivative using this connection),
$$
\Gamma^{\alpha}_{. \mu \nu} \edf \h{i}^{\alpha} \h{i}_{\mu , \nu}
= \cs{\mu}{\nu}{\alpha} + \gamma^{\alpha}_{ . \mu \nu},
\eqno{(10)}
$$
where $\cs{\mu}{\nu}{\alpha}$ is Christoffel symbol of second the
second kind and $\gamma^{\alpha}_{. \mu \nu}$ is the contortion
tensor, given by
$$
\gamma^{\alpha}_{. \mu \nu} \edf \h{i}^{\alpha} \h{i}_{\mu ; \nu},
\eqno{(11)}
$$
the semicolon is used to characterize covariant differentiation
using Christoffel symbol. Now the torsion of the AP-space is
defined by
$$
\Lambda^{\alpha}_{. \mu \nu} = \Gamma^{\alpha}_{. \mu \nu} -
\Gamma^{\alpha}_{. \nu \mu} = \gamma^{\alpha}_{. \mu \nu} -
\gamma^{\alpha}_{. \nu \mu }. \eqno{(12)}
$$
The relation between (11) and (12) can be written as$^{7}$,
$$
\gamma^{\alpha}_{\mu \nu} = \frac{1}{2} (\Lambda^{\alpha}_{.\mu
\nu}- \Lambda^{~\alpha~~}_{\mu . \nu}- \Lambda^{~\alpha~}_{\nu .
\mu}) \eqno{(13)}
$$
The relations(12) and (13) implies that the vanishing of the
torsion is a necessary and sufficient condition for the vanishing
of the contortion. The curvature tensor corresponding to the
linear connection (10) can be written, in the usual manner, as
$$
B^{\alpha}_{~\mu \nu \sigma} \edf \Gamma^{\alpha}_{~ \mu \sigma ,
\nu}- \Gamma^{\alpha}_{~ \mu \nu , \sigma} +
\Gamma^{\epsilon}_{~\mu \sigma}\Gamma^{\alpha}_{~\epsilon
\nu}-\Gamma^{\epsilon}_{~\mu \nu}\Gamma^{\alpha}_{~\epsilon
\sigma} . \eqno{(14)}
$$
This tensor vanishes identically because of (9). Although this
appears as a disappointing feature, but we are going, in the next
section, to show that it represents a cornerstone in the present
work. The curvature tensor (14) can be written in the form,
$$
B^{\alpha}_{~\mu \nu \sigma}= R^{\alpha}_{~\mu \nu \sigma}+
Q^{\alpha}_{~\mu \nu \sigma} \equiv 0,   \eqno{(15)}
$$
where,
$$
R^{\alpha}_{~\mu \nu \sigma} \edf \cs{\mu}{\sigma}{\alpha}_{, \nu}
- \cs{\mu}{\nu}{\alpha}_{, \sigma}+
\cs{\mu}{\sigma}{\epsilon}\cs{\epsilon}{\nu}{\alpha}
-\cs{\mu}{\nu}{\epsilon}\cs{\epsilon}{\sigma}{\alpha}, \eqno{(16)}
$$
is  the Riemann- Christoffel curvature tensor and the tensor
$Q^{\alpha}_{\mu \nu \sigma}$ is defined by
$$
Q^{\alpha}_{~\mu \nu \sigma} \edf \gamma^{\stackrel{\alpha}{+}}_{~
\stackrel{\mu}{+}\stackrel{\sigma}{+}|\nu}-\gamma^{\stackrel{\alpha}{+}}_{~
\stackrel{\mu}{+}\stackrel{\nu}{-}|\sigma}- \gamma^{\beta}_{~ \mu
\sigma}\gamma^{\alpha}_{\beta \nu} +  \gamma^{\beta}_{~ \mu
\nu}\gamma^{\alpha}_{\beta \sigma}, \eqno{(17)}
$$
where the (-)sign is used to characterize covariant derivatives
using the dual connection $\tilde \Gamma^{\alpha}_{. \mu \nu} (=
\Gamma^{\alpha}_{.\nu \mu} )$ .

Now to get a non-vanishing curvature, we have to parameterize (10)
by replacing it with$^{8}$,
$$
\nabla^{\alpha}_{. \mu \nu} = \cs{\mu}{\nu}{\alpha} + b ~
\gamma^{\alpha}_{. \mu \nu}, \eqno{(18)}
$$
where $b$ is a dimensionless parameter. The curvature tensor
corresponding to (18) can be written as,
$$
{\hat{B}}^{\alpha}_{.\mu \nu \sigma}= R^{\alpha}_{.\mu \nu
\sigma}+ b~ Q^{\alpha}_{.\mu \nu \sigma},   \eqno{(19)}
$$
which is, in general, a non-vanishing tensor. The version of the
AP-geometry built using (18) is known as the PAP-geometry$^{9}$.
The path equations corresponding to (18) can be written as,
$$
\frac{d^{2}x^{\mu}}{d\tau^{2}} +
\cs{\alpha}{\beta}{\mu}\frac{dx^{\alpha}}{d\tau}
\frac{dx^{\beta}}{d\tau} = - b~~\Lambda^{..\mu}_{\alpha
\beta.}\frac{dx^{\alpha}}{d\tau} \frac{dx^{\beta}}{d\tau} .
\eqno{(20)}
$$
where $\tau$ is a scalar parameter. It can be easily shown that
(18) defines a non-symmetric, metric linear connection.
\section{Torsion Energy and Energy Densities}
Using equation (15) we can write
$$
 R^{\alpha}_{.\mu \nu \sigma}
\equiv - Q^{\alpha}_{.\mu \nu \sigma}.   \eqno{(21)}
$$
Although the tensors $R^{\alpha}_{. \mu \nu \sigma}$ and
$Q^{\alpha}_{. \mu \nu \sigma}$ appear to be mathematically
equivalent, they have the following differences:\\
1- The Riemann-Christoffel curvature tensor is made purely from
Christoffel symbols

(see (16)), while the tensor $Q^{\alpha}_{.\mu \nu \sigma}$ (17)
is made purely from the contortion (11)

( or from the torsion using (13)). The first tensor is
non-vanishing in Riemannian

geometry while the second vanishes in the same geometry. \\
2- The non-vanishing of $R^{\alpha}_{.\mu \nu \sigma}$ is the
measure of the curvature of the space, while the

addition of $Q^{\alpha}_{.\mu \nu \sigma}$ to it causes the space
to be flat. So, one is causing an inverse effect,

on the properties of space-time, compared to the other. For this
reason

we call $Q^{\alpha}_{.\mu \nu \sigma}$ "{\underline{\it{The
Curvature Inverse of Riemann-Christoffel Tensor}"}} $^{9}$, or

{\underline{\it "The Additive inverse of the Curvature Tensor"}}.
Note that both tensors are

considered as curvature tensor, but one of them cancels the effect
of the other, if both

existed in the same geometric structure.

Now, in view of the above two differences, we can deduce that
these two tensors are not, in general, equivalent. In other words,
if we consider gravity as curvature of space-time and is
represented by $R^{\alpha}_{. \mu \nu \sigma}$, we can consider
$Q^{\alpha}_{.\mu \nu \sigma}$ as representing
{\bf{anti-gravity}}! The existence of equal effects of gravity and
anti-gravity in the same system neutralizes the space-time,
geometrically. This situation is similar to the existence of equal
quantities of positive and negative electric charges in the same
system, which neutralizes the system electrically.

If we assume that gravity and anti-gravity effects are not exactly
equal in the same system, then space-time curvature can be
represented by the tensor (19). The existence of anti-gravity
gives rise to a repulsive force, which can be used to interpret SN
type Ia observation. This can be achieved by adjusting the
parameter $b$ . The above discussion gives an argument on the
production of a repulsive force by torsion.

It is well known that the L.H.S. of (21) satisfies the second
Bianchi Identity, so using (21) we can easily show that$^{10}$

$$
\Sigma^{\alpha \beta}_{~~; \beta} \equiv 0 \eqno{(22)}
$$
where,
$$
\Sigma^{\alpha \beta} \edf Q^{\alpha \beta}- \frac{1}{2}g^{\alpha
\beta }Q , \eqno{(23)}
$$
$$
Q_{\alpha \beta} \edf Q^{\sigma}_{~. \alpha \beta \sigma}
\eqno{(24)}
$$
$$
Q \edf g^{\alpha \beta} Q_{\alpha \beta} . \eqno{(25)}
$$
It is clear from (22) that the physical quantity represented by
the tensor $ \Sigma^{\alpha \beta}$ is a conserved quantity. We
are going to call it the {\bf{"Torsion Energy"}}, since
$Q^{\alpha}_{.\mu \nu \sigma}$ is purely made of the torsion as
mentioned above (see (17)).

As stated above, curvature and torsion energies are defined by (2)
and (23), respectively. We can take the norm of the scalers $G,
\Sigma$ to represent the curvature and torsion energy densities,
respectively. Both have the dimensions $cm^{-2}$ characterizing
energy density, in relativistic units. Now the total energy of a
system with curvature and torsion can be written as:
$$
E = \|G\| + \|\Sigma\|. \eqno{(26)}
$$
where,
$$
\Sigma \edf g_{\mu \nu} ~ \Sigma^{\mu  \nu},
$$
and
$$
G \edf g_{\mu \nu} ~ G^{\mu  \nu}.
$$
 We can define the following parameters:
$$
Curvature ~ energy ~ density ~ parameter: ~ ~ ~ ~ ~ ~ ~
\Omega_{cr} \edf \frac{\|G\|}{E}, \eqno{(27)}
$$
$$
Torsion ~ energy ~ density ~ parameter: ~ ~  ~ ~ ~ ~ \Omega_{tr}
\edf \frac{\|\Sigma\|}{E}. \eqno{(28)}
$$
Then equation (26) can be written as,
$$
1 = \Omega_{cr} + \Omega_{tr}. \eqno{(29)}
$$
If we assume that (29) represents the total energy parameter
collecting different sources of energy causing curvature and
torsion of space-time, and dark energy is due to torsion of space
time, then we can take$^{11}$ $\Omega_{tr} = \frac{2}{3}$ . This
will put a limit on the parameter $b ( \simeq 2)$ and consequently
on the parameter $\gamma (\simeq 548)$. These limits are
calculated assuming that the material constituents of the Universe
are almost fermions. Note that the value of the  parameter
$\gamma$ is unity in the case of the Earth's gravitational
field$^{12}$. It seems that its value depends on the size of the
system under consideration.

As a second argument on the existence of a repulsive force,
corresponding to the torsion of space-time, consider the
linearized form of (20) which can be written as$^{8}$,
$$
\Phi_{T} = \Phi_{N} (1-b)= \Phi_{N} + \Phi_{\Sigma}, \eqno{(30)}
$$
where,
$$
\Phi_{\Sigma} \edf - b \Phi_{N}, \eqno{(31)}
$$
 $\Phi_{N}$ is the Newtonian gravitational potential and
$\Phi_{T}$ is the total gravitational potential due the presence
of gravity and anti-gravity. It is clear from (30) that the
Newtonian potential is reduced by a factor $b$ due to the
existence of the torsion energy. It is obvious from (31) that $
\Phi_{\Sigma} $ and the Newtonian potential have opposite signs.
Then one can deduce that $\Phi_{\Sigma}$ is a {\it repulsive
gravitational potential}, in contrast to the attractive
gravitational potential $\Phi_{N}$.
\section{Discussion and Concluding Remarks}
In the present work we have chosen a version of the 4-dimensional
AP-geometry to represent the physical world including space and
time. This version, the PAP-geometry, is more wider than the
Riemannian geometry, since it has simultaneously non-vanishing
curvature and torsion. The PAP-geometry is characterized by the
parameter $b$. If $b = 0$, this geometry becomes Riemannian, while
if $b = 1$, it recovers the conventional AP-geometry.
On the present work, we can draw the following remarks:\\

1- We have applied the geometrization philosophy and its guide
lines mentioned in section 2, to a geometry with curvature and
torsion. One can summarize the results of this application in the
following
points, corresponding to the three guide lines given in section 2. \\
(i) A conservation law, as a law of nature, giving conservation of
torsion energy is represented by the differential
identity (22). \\
(ii) The quantity called the {\it "torsion energy"} is represented
by the tensor (23), which is a part of a geometric structure. This
tensor
gives rise to anti-gravity and consequently a repulsive force. \\
(iii) Trajectories of test particles affected by the torsion of
the space-time is represented  by the path equation (20), which is
a curve in the geometric structure used.

The path equation (20) has been used to study trajectories of
spinning elementary test particles in a background with torsion
and curvature. The R.H.S. of this equation is suggested to
represent a type of interaction between torsion and the quantum
spin of the moving particle$^{8}$. The application of this
equation to the motion of the thermal neutrons in the Earth's
gravitational
field removes the discrepancy in the COW-experiment$^{12}$. \\

2- It is clear that {\it torsion energy}, defined in section 4.,
can solve the problem of SN type Ia observations, since it gives
rise to a repulsive force. This can be achieved by adjusting the
parameter $b$, as shown in section 4. One can now replace the
exotic term {\it{"dark energy"}} by the term {\it{"torsion
energy"}}. The later has a known origin (a pure geometric origin).
This shows the success of the geometrization
scheme in dealing with physical problems. \\

3- Both terms {\it "dark energy"} and  {\it "cosmological
constants"} are exotic terms. They have neither  geometric origin
nor  well
defined physical origin.\\

4- Many authors consider $R^{\alpha}_{.\mu \nu \sigma}$and
$Q^{\alpha}_{.\mu \nu \sigma}$ as identical tensors from both
physical and mathematical points of view. For this reason, they
are using the term {\it{Teleparallel equivalent of GR}} for
theories built using $Q^{\alpha}_{.\mu \nu \sigma}$. In view of
the present work, such authors are, in fact, constructing
anti-gravity theories. {\it (This is similar to the situation of
constructing a field theory for positive electric charges and
another one for negative electric charges. Such theories would be
similar)}. Discrepancies would appear only if both fields existed
in the same
system, as in the case of SN type Ia observations. \\

5- Curvature and torsion corresponds to two different types of
energy. The energy corresponding to the first gives rise to an
attractive force, while the energy corresponding to the second is
repulsive. We believe that torsion energy is what has been
discovered recently by the SN type Ia observation$^{1}$. Both
energies have the same conservation law.\\

6- In the context of the present work, we can extend the
geometrization
philosophy by adding the following postulate.\\
{\underline{\it{Physical phenomenae are
just interactions between the space-time structure and}}}\\
{\underline{\it{the intrinsic properties of the material constituents of the system.}}} \\
Using this postulate, the attractive force is interpreted as a
result of the interaction between the curvature of space-time and
the mass (energy) of the material constituents; while the
repulsive force can be interpreted as a result of an interaction
between the torsion and the quantum spin of the
material constituents$^{8}$.\\

7- A successful description of gravity is to be given using a more
complete geometry, i.e. a geometry with a simultaneously
non-vanishing curvature and torsion.\\

8- The parameter $\gamma$ is a dimensionless parameter. It is
suggested to be fixed by experiment or observation$^{8}$. This
parameter characterizes the source of the background field. It has
been shown that $\gamma = 1$ in the case of the Earth as a source
of the gravitational field in the COW-experiment$^{12}$. As shown
in section 4, $\gamma = 548$ for the Universe. This parameter is
system dependent. The type of dependence of $\gamma$ can be
understood in the
context of a field theory constructed in the PAP-geometry. \\

9- The results obtained in the present work can be obtained, with
some efforts, using other geometries with curvature and torsion,
since such geometries possess similar features$^{13}$.\\ \\
{\bf Acknowledgment}\\
The author would like to thank the organizing committee of the
conference and Dr. S. Khalil for inviting him to give this talk.
\section*{References}
1. J.L. Tonry, B.P. Schmdit, et al. {\it Astrophys. J.}{\bf{594}}, 1 (2003). \\
2. P.D. Mannheim, {\it Prog.Part.Nucl.Phys.} {\bf{56}} 340 (2006); gr-qc/0505266 \\
3. S.M. Carroll, {\it"The Cosmological Constant"},
http://livingreviews.org/
Irr-2001-1. \\
4. A. Einstein, {\it "The Meaning of Relativity"}
(Princeton, 5th ed. 1955).\\
5. A. Einstein, {\it Math. Annal.} {\bf 102}, 685 (1930).\\
6. M.I. Wanas, {\it Cercet.Stiin.Ser.Mat.} {\bf 10}, 297 (2001);
gr-qc/0209050.\\
7. K. Hayashi and T. Shirafuji, T. {\it Phys.Rev.} D{\bf 19},
3524 (1979). \\
8. M.I. Wanas, {\it Astrophys. Space Sci.}, {\bf{258}}, 237 (1998); gr-qc/9904019 \\
9. M.I.Wanas, {\it Turk. J. Phys.} {\bf{24}}, 473 (2000); gr-qc/0010099 \\
10. M.I. Wanas, arXiv:0705.2255 (2007). \\
11. W.L.Friedman and M.S. Turner, {\it Rev.Mod.Phys.} {\bf 75} 1433 (2003). \\
12. M.I.Wanas, M. Melek and M.E. Kahil,  {\it Gravit.
Cosmol.},{\bf{6}}, 319 (2000).\\
13. M.I. Wanas and M.E. Kahil, {\it Gen.Rel.Gravit.} {\bf 31} 1921 (1999); arXiv:9912007. \\
\end{document}